\documentclass[aip,amsmath,amssymb,reprint]{revtex4-1}

\usepackage{bm}
\usepackage{amsmath}
\usepackage{graphicx}
\usepackage{lineno}
\usepackage{etoolbox}
\usepackage[utf8]{inputenc}
\usepackage[T1]{fontenc}
\usepackage{mathptmx}
\usepackage{dcolumn}

\begin{document}

\preprint{AIP/123-QED}

\title[Cooperative near- and far-field thermal management via diffusive superimposed dipoles]{Cooperative near- and far-field thermal management via diffusive superimposed dipoles}

\author{Pengfei Zhuang}
\affiliation{Department of Physics, State Key Laboratory of Surface Physics, and Key Laboratory of Micro and Nano Photonic Structures (MOE), Fudan University}


\author{Xinchen Zhou}
\affiliation{Department of Physics, State Key Laboratory of Surface Physics, and Key Laboratory of Micro and Nano Photonic Structures (MOE), Fudan University}

\author{Liujun Xu}
\email{ljxu@gscaep.ac.cn}
\affiliation{Graduate School of China Academy of Engineering Physics}

\author{Jiping Huang}
\email{jphuang@fudan.edu.cn}
\affiliation{Department of Physics, State Key Laboratory of Surface Physics, and Key Laboratory of Micro and Nano Photonic Structures (MOE), Fudan University}

\begin{abstract}
Active metadevices with external excitations exhibit significant potential for advanced heat regulation. Nonetheless, conventional inputs, like heating/cooling and introducing convection by rotating plate, display inherent limitations. One is the only focus on far-field control to eliminate temperature distortion in the background while neglecting near-field regulation in the functional region. Another is lacking adaptability due to complex devices like thermoelectric modules and stepping motors. To tackle these challenges, the concept of diffusive superimposed dipoles characterized by orthogonal thermal dipole moments is proposed. Cooperative near- and far-field regulation of temperature fields is achieved by designing superimposed dipole moments, enabling transparency and cloaking functionalities. Simulation and experiment outcomes affirm the efficacy of this adaptive thermal field control technique, even when interface thermal resistance is taken into account. Adaptivity stems from dipole moment decomposability, allowing metadevices to operate in various heat flux directions and background thermal conductivity. These findings could pave the way for cooperative and adaptive thermal management and hold potential applications in other Laplace fields, including direct current and hydrodynamics.
\end{abstract}

\maketitle
\date{\today}
\section{Introduction}\label{1}

Efficient and precise temperature field control has been a long-standing subject of interest in the field of thermal management~\cite{SongJoule18,ZengAE20,ZhuAIPAdvance15}. Nowadays, thermal management primarily entails the process of modulating and controlling temperature and temperature gradients through heating or cooling means. For instance, to ensure the efficient operation of various devices like chips~\cite{MathewJEP22,HeAE22} and batteries~\cite{LongchampsACSEL22,YueAE24}, it is imperative to maintain them within suitable temperature ranges. From an integrated systems standpoint, when endeavors are undertaken to regulate localized temperature distributions via targeted heating and cooling, it becomes essential to concurrently maintain the integrity and functionality of neighboring components within the expansive system framework. Recently, diverse thermal metamaterials~\cite{ZhangNRP23,YangPR21,RanAM23,YangRMP23,XuIJHMT21,XuPRAP19b,XuPRE18,ShenAPL16}, functional building blocks densely packed into an effective material, have been designed to achieve advanced functionalities, including thermal cloak~\cite{XuPRL14b,DaiJAP18,XuCPL20}, concentrator~\cite{HanEES13,YangESEE19}, and thermal camouflage~\cite{XuECM19,HuMT21}. Depending on whether external inputs are required, thermal metamaterials could be generally classified into passive and active schemes. 

\begin{figure*}
	\centering
		\includegraphics[width=\linewidth]{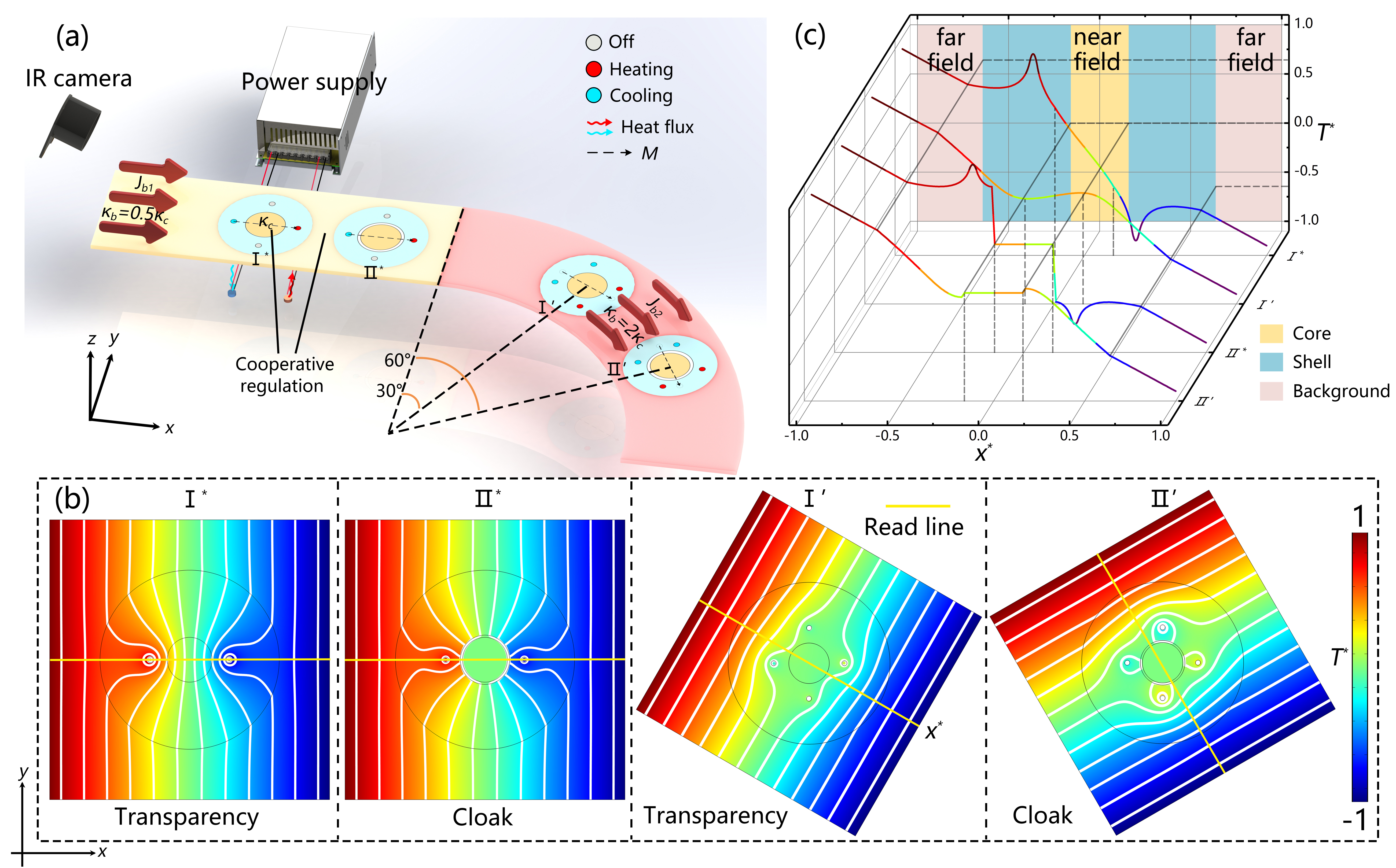}
	  \caption{Principle of adaptive metadevices regulated by diffusive superimposed dipoles. (a) Schematic of superimposed-dipole metadevices: transparency, and cloak, applied in different scenarios. (b) The steady-state numerical temperature fields for schemes \uppercase\expandafter{\romannumeral1}$^{*}$-\uppercase\expandafter{\romannumeral2}$^{\prime}$. (c) The middle-line temperature distributions of transparency and cloak in different scenarios.}\label{fig1}
\end{figure*}

Representative passive frameworks include transformation theory, scattering cancellation theory, and effective medium theory. Inspired by the concepts of Albert Einstein’s theory of general relativity, the transformation theory~\cite{PendryScience06,LeonhardtScience06,XuESEE20} originates from transformation optics and has been extended to many other physical fields, such as acoustics~\cite{MaSa16}, thermotics~\cite{FanAPL08a,LiNRM21}, particle dynamics~\cite{GuenneauJSI13}, and plasma physics~\cite{ZhangCPL22}. The powerful theoretical framework provides guidance for the parameters we need to manipulate the thermal field at will. However, the cost is that the required parameters, such as thermal conductivity, are anisotropic and inhomogeneous. This is closely related to the non-conformality and inhomogeneity of the coordinate transformation. Recent studies have used pseudo-conformal transformations~\cite{XuNP15,DaiCSF23} and linear transformations~\cite{LiuIJHMT17,XuECM18} to eliminate the anisotropy and inhomogeneity of material parameters, respectively, which greatly reduces the difficulty of experimental preparation. Alternative theory, like scattering cancellation~\cite{HanPRL14,MaPRL14} have been utilize to fabricate thermal metamaterials with the desired properties. Because this strategy is implemented by directly solving the Laplace equation, it can only analytically solve the temperature fields of geometrically isotropic structures with anisotropic material parameters~\cite{JinIJHMT20} or geometrically anisotropic structures with isotropic material parameters~\cite{YangAPL17a}. For complex geometric structures~\cite{XuJAP18} with anisotropic material parameters, a set of parameter configurations can be proposed to eliminate the disturbance of the background temperature field, but the entire temperature field cannot be precisely solved. The effective medium theory~\cite{NarayanaPRL12,DongJAP04,HuangAPL05} is a common approach that can achieve equivalent complex properties derived from transformation thermotics or scattering cancellation theory, where the Maxwell–Garnett theory and the Bruggeman formula are two basic theoretical tools. However, these traditional formulas often fail in complex geometric configurations and at high volume fractions. Fortunately, due to the development of computer science, the effective medium theory is gradually being combined with optimization algorithms~\cite{ShaNC21} or machine learning~\cite{HuPRX20}, which greatly improves computational efficiency and accuracy. To summarize the aforementioned methods of passive thermal field manipulation, they often require carefully designing the spatial distribution of materials. This means that when background parameters change, the configuration of the materials also needs to be adjusted.

In contrast, active schemes utilize external excitations to achieve impressive abilities of thermal manipulation without redesigning their architectures. Manual temperature control enables intriguing functionalities like thermal invisibility in complex scenarios~\cite{NguyenAPL15,LiNC22,YePASMAP08,LiuPO13}, thermal convection driven by external forces helps enhance heat transfer~\cite{JinPNAS23,LiNM19}, and spatiotemporal modulation~\cite{XuPRL22a,XuPRL22b} provides an unexpected method to realize asymmetric heat transfer. These active schemes significantly improve heat control ability and diversity. Despite powerful control of temperature fields, some common restrictions of active schemes remain to be solved. On the one hand, though far-field temperature control by continuous boundary setting~\cite{LiNC22,NguyenAPL15} or thermal dipoles~\cite{XuPRE19a} offsets the disturbance of inclusions in the background, near-field regulation in the functional region is often neglected, constraining the degree of freedom in regulating thermal fields. On the other hand, the complex devices for active control often include thermoelectric modules or stepping motors, lacking adaptability to environmental changes, such as heat flux directions and background thermal conductivity. These challenging problems severely hinder cooperative and adaptive heat control.

To address these challenges, the diffusive superimposed dipoles are proposed and applied to a core-shell structure [Fig.~\ref{fig1}(a)]. The superimposed dipoles are constructed from multiple dipole pairs and each dipole consists of a heat source and a cold source. The power of the heat source and cold source is designed artificially, and can be achieved through methods such as semiconductor heating/cooling pieces controlled by power supply, or direct thermal contact with a constant temperature water bath. Building on this concept, thermal transparencies [see schemes I$^*$ in Fig.~\ref{fig1}(a)] with tunable central heat flux are designed and bilayer-like structures with thin insulating inner shells are demonstrated to realize thermal cloaking [see schemes II$^*$ in Fig.~\ref{fig1}(a)]. All schemes have experienced a change in scenarios, that is, transitioning from a yellow zone (superscript $^*$) to a pink zone (superscript $^{\prime}$), where both the direction of environmental heat flow and the environmental thermal conductivity change. It is clearly observable from the temperature profile [Fig.~\ref{fig1}(b)] and middle-line temperature distribution [Fig.~\ref{fig1}(c)] that these devices can effectively fulfill their intended functions before and after the scenario transition. The significances of this work are fourfold: (I) Unlike traditional active methods primarily focusing on far-field control, our scheme yields cooperative near- and far-field (core and background region) regulation by diffusive superimposed dipoles, facilitating full control of heat flow; (II) Differing from traditional heat management methods of direct heating/cooling, superimposed dipoles can adjust the near-field temperature gradient without changing the far-field temperature gradient, preventing interference with the normal operation of surrounding devices; (III) The proposed metadevices adapt to varying background thermal conductivity and heat flux direction without modifying inherent parameters, enhancing engineering practicality; (IV) These schemes are achieved with isotropic and homogeneous materials, contributing to simple device preparation.


\section{Scheme design and numerical validation}\label{2}
\subsection{Circular meta-devices resorting to diffusion-superposition dipoles}
\begin{figure}
	\includegraphics[width=\linewidth]{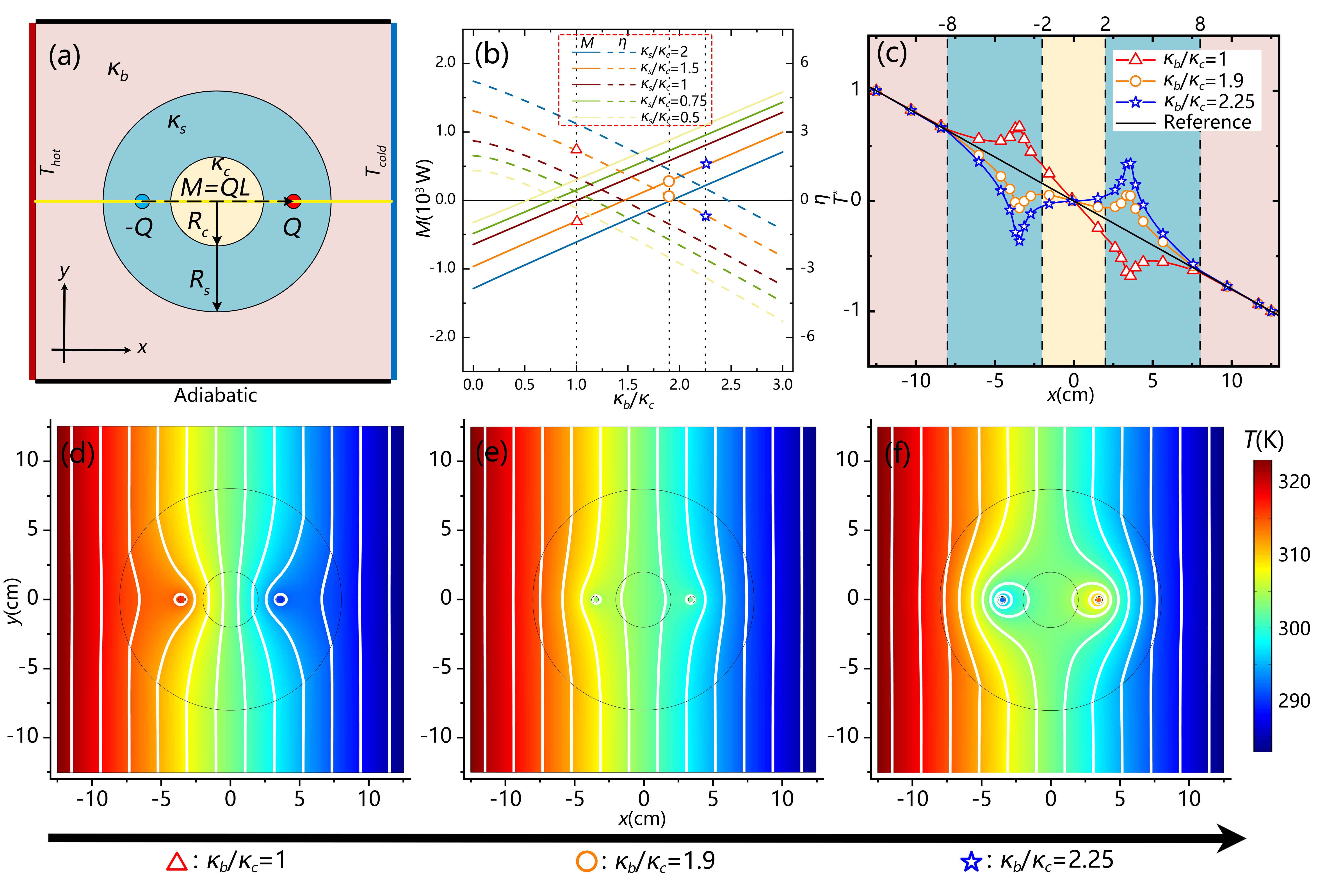}
	\caption{Design and simulation demonstrations of dipole-induced transparency. (a) Schematic diagram of dipole-induced transparency. (b) Concentrating efficiency $\eta$ and dipole moment $M$ as a function of $\kappa_b/\kappa_c$ with a constant $\kappa_s/\kappa_c$. The temperature distribution of the middle line in (a) and cross-section are plotted in (c) and (d)-(f), respectively.}
	\label{fig2}
\end{figure}

Two-dimensional heat conduction in a homogeneous and isotropic medium is considered, where a circular core-shell structure is embedded in a finite background with thermal conductivity $\kappa_b$ [Fig.~\ref{fig2}(a)]. The radii of the core and shell are denoted by $R_c$ and $R_s$, respectively, with their thermal conductivities being $\kappa_c$ and $\kappa_s$. While traditional passive metadevices often adjust the thermal conductivity of the shell, they exhibit a lack of adaptability to changing backgrounds. To enhance flexibility, diffusive superimposed dipoles are introduced into the shell based on specific requirements. For the sake of simplicity, a fixed dipole moment $M=QL$ and an external temperature gradient $\nabla T_b$ in the $x$ direction are initially considered. Here, $Q$ and $L$ denote the power and distance of the thermal dipole, respectively. Note that the radius of the monopole needs to be much smaller than the distance between the dipoles. The steady-state heat conduction equation in the polar coordinate system $(r,\theta)$ can be expressed as:
\begin{equation}
	\kappa\left(\frac{\partial^{2} T}{\partial r^{2}}+\frac{1}{r} \frac{\partial T}{\partial r}+\frac{1}{r^{2}} \frac{\partial^{2} T}{\partial \theta^{2}}\right)=0.
\end{equation}

Due to the linearity of the Laplace equation, the temperature field distributions of the external source and the thermal dipole can be computed separately and then combined. In the presence of only a uniform temperature gradient, the temperature distribution is described by a fundamental solution $T(r,\theta)=(Ar+B/r)\cos \theta$. This leads to the following temperature profiles:
\begin{equation}\label{Te2}
	\left\{\begin{aligned}
		T_{e,c}&=A_{c} r \cos \theta +T_0 \left(r<R_{c}\right),\\
		T_{e,s}&=A_{s} r \cos \theta+B_{s} r^{-1} \cos \theta+T_0 \left(R_{c}<r<R_{s}\right), \\
		T_{e,b}&=A_{b} r \cos \theta+B_{b} r^{-1} \cos \theta +T_0 \left(r>R_{s}\right),
	\end{aligned}\right.
\end{equation}
where $T_0$ is the reference temperature and the coefficients $A_c, A_s, B_s, B_b$, and $A_b$ can be determined by the boundary conditions (see Supporting Information Note 1 for detailed calculations). It should be noted that when a central inclusion exists, the background temperature tends to be disturbed, indicating that $B_b\ne 0$.
Then, the temperature profile solely induced by the thermal dipole is calculated. The general temperature distribution resulting from the thermal dipole is expressed as:~\cite{XuPRE19a}
\begin{equation}
	T_d=\frac{-Q}{2 \pi \kappa} \ln r_{+}+\frac{Q}{2 \pi \kappa} \ln r_{-},
\end{equation}
where $r_{+}$ and $r_{-}$ represent the distances to the heater and cooler of the dipole, respectively. In different regions, the temperature field can be approximated as follows: $T_{d,c}\approx 2 M r \cos \theta/(\pi \kappa_c L^{2})$ in the near field $(r \ll L)$, $T_{d,b} \approx D_{b} \cos \theta /r$ in the far field $(r \gg L)$, and $T_{d,s} \approx C_s r \cos \theta+M \cos \theta/(2 \pi \kappa_s r)$ in the middle field $(r \sim L)$ (see Supporting Information Note 1 for detailed calculations). Since the dipole is placed in the shell, its near field, middle field, and far field regions almost coincide with the core, shell, and background regions of the mixed structure, respectively. Interestingly, the expressions for the near-field and far-field temperatures are similar to Eq.~(\ref{Te2}). Then, the total thermal field, considering the combined effect of the external heat source and thermal dipole, is derived as:
\begin{equation}\label{Tt2}
	\left\{\begin{aligned}
		T_{t,c}&=\nabla T_{c}r \cos \theta+T_0,\\
		T_{t,b}&=\nabla T_{b}r\cos \theta+(B_s+D_s)r^{-1}\cos \theta+T_0,
	\end{aligned}\right.
\end{equation}
where $\nabla T_{c}$=$2M/(\pi \kappa_c L^2)+A_c$. Our objective is to regulate the temperature in the core without disturbing the background (transparency) by utilizing the near-field and far-field thermal dipole effects. To fully address this problem, the value of \( M \) is determined by solving the equation \( B_s+D_s=0 \) (see Supporting Information Note 1 for details). Variations of the dipole moment \( M \) and concentrating efficiency \( \eta=\nabla T_{c}/\nabla T_{b} \) are depicted in [Fig.~\ref{fig2}(b)], treated as functions of \( \kappa_b/\kappa_c \) while maintaining \( \kappa_s/\kappa_c \) constant. The dipole moment and concentrating efficiency are plotted for $\kappa_b/\kappa_c$=1 (red triangle), $\kappa_b/\kappa_c$=1.9 (orange circle), and $\kappa_b/\kappa_c$=2.25 (blue star) by setting $\kappa_s/\kappa_c$=1.5. According to the theoretical predictions shown in Fig.~\ref{fig2}(b), the dipole moment and shell conductivity can be tailored to achieve thermal transparency with desired local temperature gradients. If the thermal conductivity of the background changes, the thermal fields of both the background and the central region will be disturbed. At this time, the dipole moment can be adjusted as a flexible parameter to offset the disturbance of the background thermal field.

Simulative demonstrations of dipole-induced transparency [Fig.~\ref{fig2}(a)] are conducted using COMSOL Multiphysics to validate the proposed designs. Taking into account the practical fabrication, the size of the background is set at \( 25\times 25 \)~cm\(^2\), with the core and shell radii defined as \( R_c=2 \)~cm and \( R_s=8 \)~cm, and the dipole distance at \( L=7 \)~cm. A temperature gradient is applied in the background by maintaining $T_{hot}$=323~K and $T_{cold}$=283~K. The other boundaries are adiabatic. The yellow line delineated along the central axis is employed to extract simulation results, which are subsequently plotted as curves depicted in Fig.~\ref{fig2}(c). Numerical simulations of the temperature profiles for transparency at three distinct points are illustrated in Figs.~\ref{fig2}(d)-\ref{fig2}(f) using a core thermal conductivity \( \kappa_c=200 \)~W~m\(^{-1}\)~K\(^{-1}\) and shell thermal conductivity \( \kappa_s=1.5\kappa_c \). The disturbance in the isotherms caused by the thermal conductivity mismatch is effectively canceled out by the thermal field generated by the thermal dipole. Simultaneously, the dipole moment should be adjusted as the background thermal conductivity changes, influencing the temperature gradient near the center. The validity of these findings is further corroborated by the degree of fit of the curve observed in the background region of Fig.~\ref{fig2}(c). A dimensionless temperature $T^*$=$(T-T_0)/(T_L-T_R)$ is introduced for convenience. The outcomes reveal that, the temperature gradient in the center is enhanced ($\eta>1$) with $\kappa_b/\kappa_c$=1 [Fig.~\ref{fig2}(d)], weakened ($\eta$<1) with $\kappa_b/\kappa_c=1.9$ [Fig.~\ref{fig2}(e)], and even reversed ($\eta$<0) with $\kappa_b/\kappa_c$=2.25 [Fig.~\ref{fig2}(f)] as expected. Such findings confirm that the proposed design effectively enables thermal transparency and regulates the temperature gradient near the center, by varying the dipole moment in relation to the background thermal conductivity.

\begin{figure}
	\includegraphics[width=\linewidth]{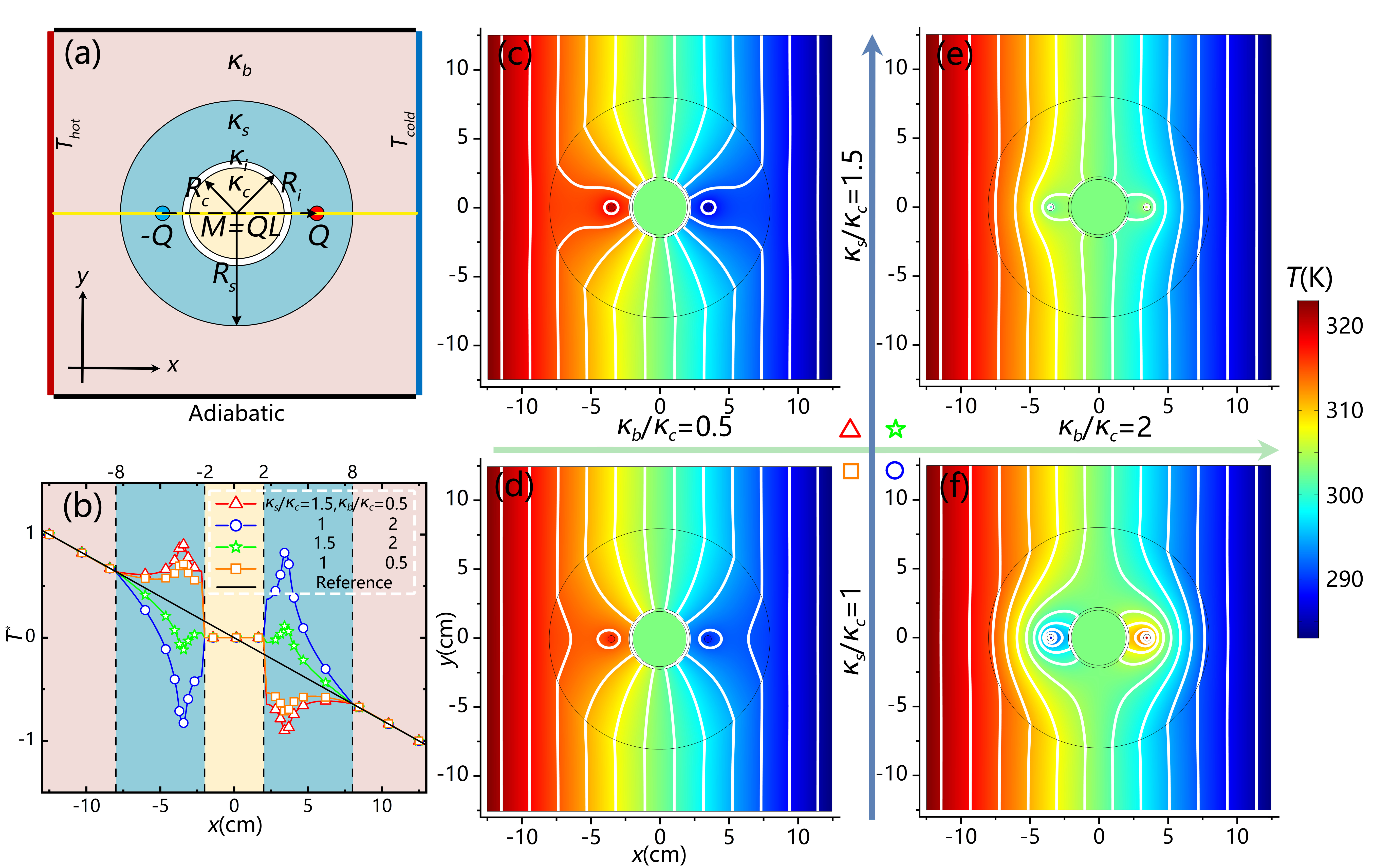}
	\caption{Design and simulation demonstrations of dipole-induced cloak. (a) Schematic of dipole-induced cloaks demonstrated with $\kappa_s/\kappa_c$=1.5 and $\kappa_s/\kappa_c$=1. (b) Middle-line $T^{*}$ of these two schemes in different background thermal conductivities. Temperature profiles of scheme (c)(e) 1 and (d)(f) 2.}
	\label{fig3}
\end{figure}

Here, the design is extended to an adaptive thermal cloak capable of protecting an object from external heat flux and remaining invisible to infrared cameras. Drawing inspiration from the bilayer structure, a thin insulating layer [Fig.~\ref{fig3}(a)] is introduced between the core and the outer shell, allowing the isotherms to bypass the core perfectly. By manipulating the dipole moment, a uniform temperature gradient outside the metadevice is achieved. Two cloaks are constructed with an insulating layer thickness $R_i-R_c=0.2$~cm and the other geometric parameters are the same as those in Fig.~\ref{fig2}. Utilizing the yellow line in Fig.~\ref{fig3}(a), the central-axis temperature distribution is extracted and illustrated in Fig.~\ref{fig3}(c). Figs.~\ref{fig3}(c) and~\ref{fig3}(e) display the temperature profiles of the first cloak ($\kappa_s/\kappa_c=1.5$) in various background thermal conductivities $\kappa_b/\kappa_c=0.5$ and $\kappa_b/\kappa_c=2$, respectively. The simulative results of the second cloak ($\kappa_s/\kappa_c=1$) are presented in Figs.~\ref{fig3}(d) and~\ref{fig3}(f). Clearly, the isotherms outside the shells are straight as if there are no core-shell structures, and the temperatures in the cores remain constant as expected. To quantitatively evaluate such performances, the central-axis temperature distribution of Figs.~\ref{fig3}(c)-\ref{fig3}(f) are recorded in Fig.~\ref{fig3}(b). These data reveal that the measured curves of cloaks overlap well with the reference line (black solid line) and the temperature gradient in the central region is approximately zero. Through the near-field and far-field thermal dipole effects, the thermal transparency and cloak can adapt to different thermal conductivities without redesigning their architectures.

The initial discussions focused on the background heat flow in the $x$-axis direction. However, practical application scenarios are complex and require metadevices that can handle heat flux from any direction. To achieve this, two pairs of thermal dipoles are placed along the $x$ and $y$ direction, with a general heat flux $J_{B}$ rotating by an angle $\theta_0$. The relationship between the component dipole moments $M_1$ and $M_2$, and the effective dipole moment $M$, follows a vector decomposition-like relationship (see Supporting Information Note 2 for detailed derivations),
\begin{equation}\label{relation_M}
	\left\{\begin{aligned}
		M_1&=M \cos \theta_0, \\
		M_2&=M \sin \theta_0.
	\end{aligned}\right.
\end{equation}
\begin{figure}
	\includegraphics[width=\linewidth]{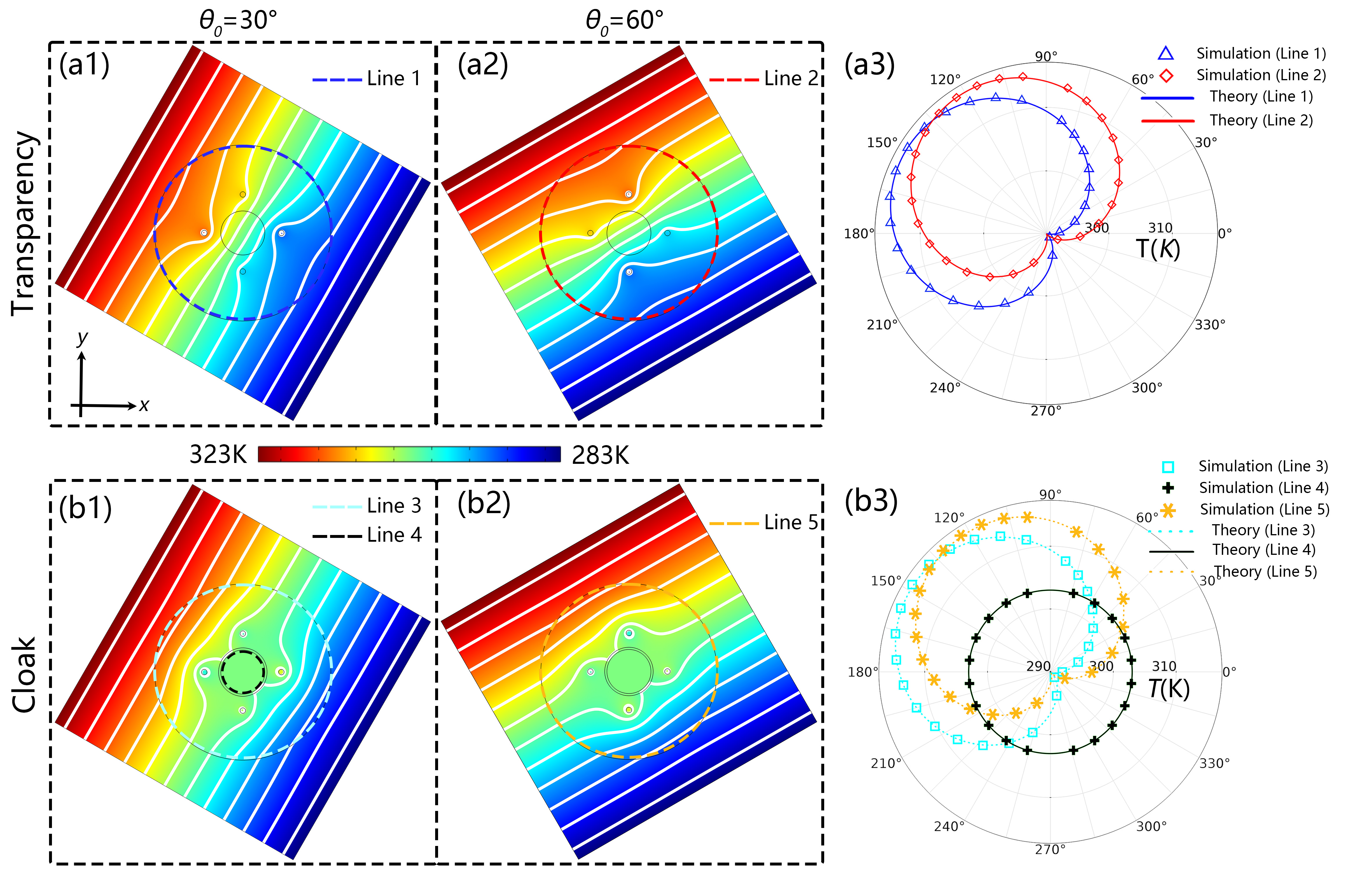}
	\caption{Validation of Dipole superposition effect. Temperature profiles of (a1)(a2) transparency rotated from Fig.~\ref{fig2}(d) and (b1)(b2) cloak rotated from Fig.~\ref{fig3}(e). (a3)(b3) Temperature distribution along the boundaries of the corresponding color in (a1)-(b2).}
	\label{fig4}
\end{figure}
Using Eq.~(\ref{relation_M}), the dipole moment can be omnidirectionally controlled by adjusting the powers of dipoles, eliminating the need for any structural modifications. This flexibility indicates significant potential for practical implementations. It is worth noting that if we analogize the potential distribution of the electric quadrupole, these two pairs of thermal dipoles are essentially equivalent to a single dipole. To validate this general case, simulations are conducted on the metadevices with geometric parameters and thermal conductivity configurations identical to those in Figs.~\ref{fig2}(d) and~\ref{fig3}(e). Figs.~\ref{fig4}(a1), \ref{fig4}(a2) and~\ref{fig4}(b1), \ref{fig4}(b2) show the temperature profiles of transparency and cloak under rotating angles of $30^{\circ}$ and $60^{\circ}$, respectively. The temperature variations as a function of the angle along different boundaries are presented in Figs.~\ref{fig4}(a3) and~\ref{fig4}(b3), where the scattered points represent the simulated results, and the continuous lines indicate the theoretical predictions. Apparently, the simulated temperature overlaps well with the form of Eq.~(\ref{Tt2}), validating the accuracy of the dipole approximation.

\subsection{Elliptical meta-devices resorting to diffusion-superposition dipoles}

\begin{figure}
\centering
	\includegraphics[width=\linewidth]{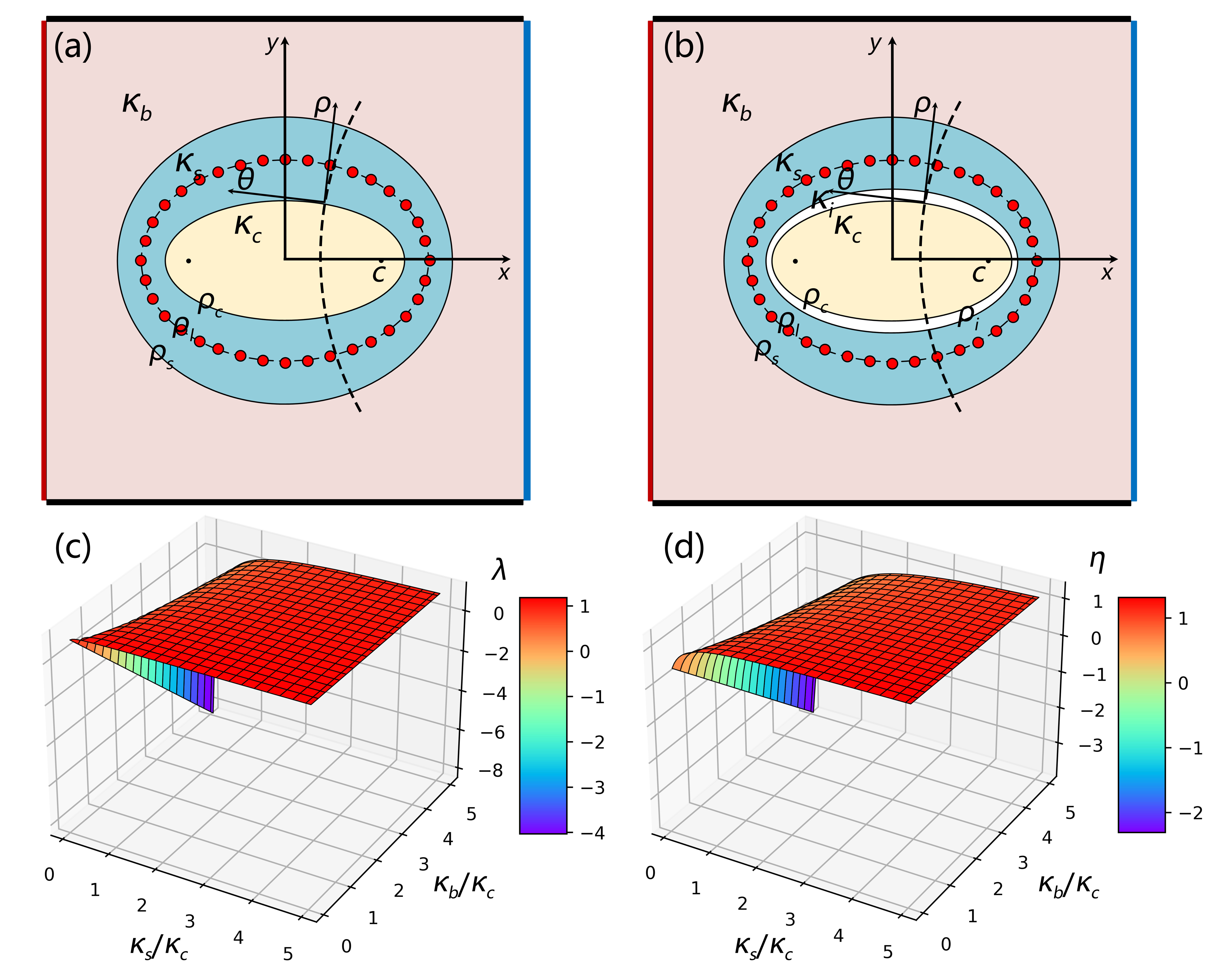}
	\caption{Schematic diagram of elliptical thermal (a) transparency and (b) cloak induced by superimposed dipoles (equivalent to line boundary heat source). (c) Gradient ratio $\lambda$=$\nabla T_l/ \nabla T_u$ of the transparency and cloak as a function of $\kappa_s/\kappa_c$ and $\kappa_b/\kappa_c$. (d) Concentrating efficiency $\eta$=$\nabla T_c/ \nabla T_u$ of the transparency as a function of $\kappa_s/\kappa_c$ and $\kappa_b/\kappa_c$.}
	\label{fig5}
\end{figure}

In this section, we consider a confocal elliptical core-shell structure embedded in a finite background, as shown in Figs.~\ref{fig5}(a) and \ref{fig5}(b). The corresponding curvilinear system is the elliptic coordinate system ($\rho,\theta$), which is related to the Cartesian coordinates as follows:
\begin{equation}\label{e_coordinate}
\left\{\begin{aligned}
x&=c\cosh(\rho) \cos(\theta),\\
y&=c\sinh(\rho) \sin(\theta).
\end{aligned}\right.
\end{equation}
The coordinate $\rho$ plays the role that the radius plays in spherical coordinates. For example, the boundaries of core, line heat source, and shell are respectively denoted as $\rho_c$, $\rho_l$, and $\rho_s$. In the presence of a line boundary heat source $x\nabla T_l$ and an external heat source $x\nabla T_u$, the heat conduction equation can be expressed as:
\begin{equation}\label{Laplace_e}
\frac{\partial^{2} T}{\partial \rho^{2}}+\frac{\partial^{2} T}{\partial \theta^{2}}=0.
\end{equation}

The temperature distribution of each region is given by~\cite{LiNC22}:
\begin{equation}\label{Tte}
\left\{\begin{aligned}
T_{c}&=E_c \cosh(\rho) \cos(\theta), \\
T_{s 1}&=\left[E_{s 1} \cosh(\rho)+F_{s 1} \sinh(\rho)\right] \cos(\theta), \\
T_{s 2}&=\left[E_{s 2} \cosh(\rho)+F_{s 2} \sinh(\rho)\right] \cos(\theta), \\
T_{b}&=\left[E_b \cosh(\rho)+F_b \, e^{-\rho}\right] \cos(\theta),
\end{aligned}\right.
\end{equation}
where the shell temperature is divided into $T_{s 1}$ and $T_{s 2}$ due to the constant temperature boundary condition. By setting $F_b=0$ and applying the outer boundary conditions, these seven coefficients can be determined (See Supplementary Information Note 3 for detailed calculation). For a comprehensive characterization of the temperature control of the core and background, gradient ratio $\lambda$ and concentrating efficiency $\eta$ are defined:
\begin{widetext}
\begin{equation}\label{le}
\left\{\begin{aligned}
\lambda&=\frac{\nabla T_l}{\nabla T_u}=\left[\kappa_s \cosh^2(\rho_s)-\kappa_b \sinh^2(\rho_s)+(\kappa_b-\kappa_s)\cosh(\rho_s) \sinh(\rho_s) \tanh(\rho_l)\right]/\kappa_s,\\
\eta&=\frac{\nabla T_c}{\nabla T_u}=\frac{(\kappa_b+\kappa_s)\cosh(\rho_l)+(\kappa_s-\kappa_b) \cosh (\rho_l-2\rho_s)}{(\kappa_s-\kappa_c) \cosh (2\rho_c-\rho_l) +(\kappa_c+\kappa_s) \cosh(\rho_l)}.
\end{aligned}\right.
\end{equation}
\end{widetext}

\begin{figure}
\centering
	\includegraphics[width=\linewidth]{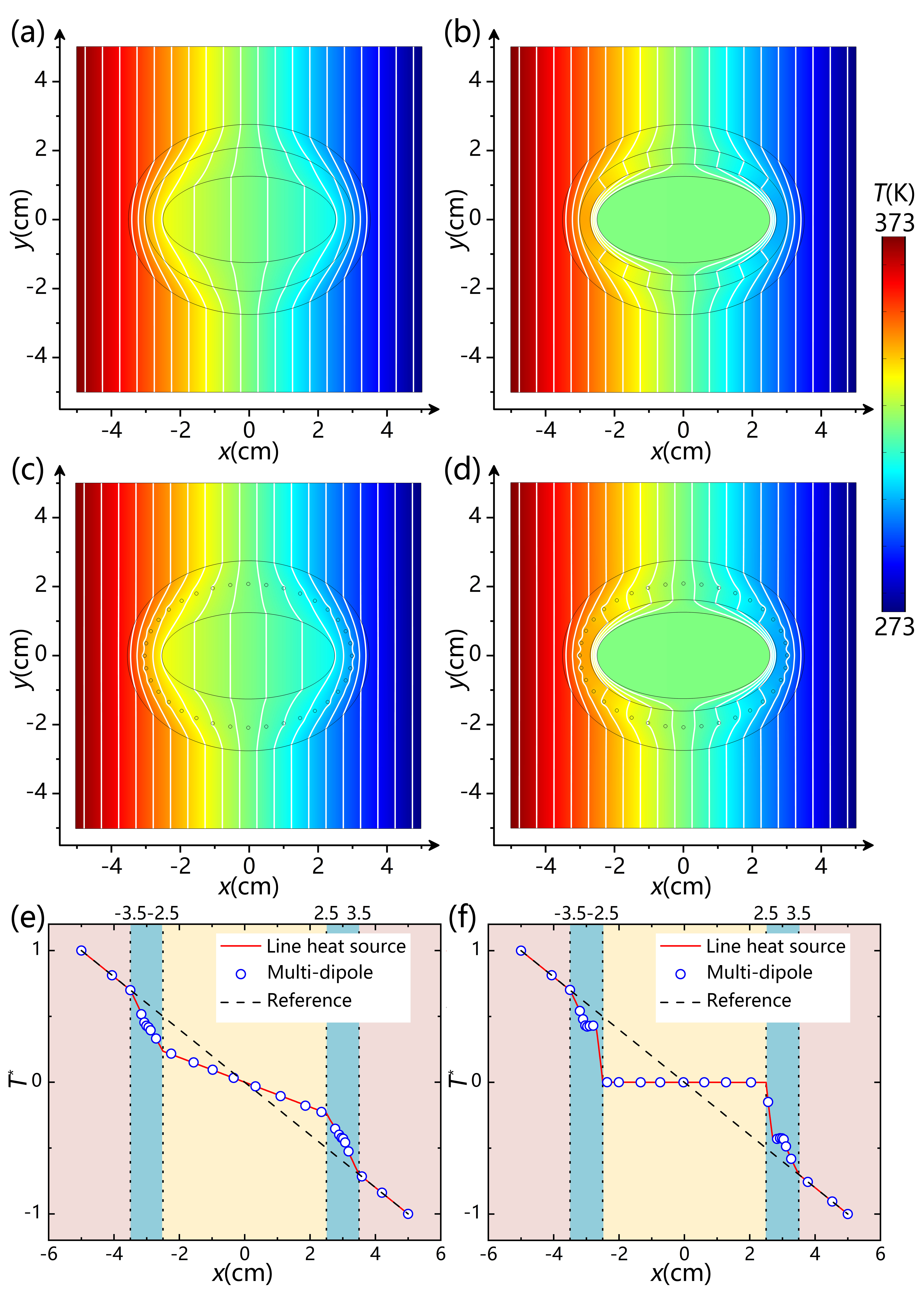}
	\caption{Superimposed dipoles for confocal elliptical structure. Temperature profiles of thermal (a) transparency and (b) cloak assisted by a line boundary heat source. In (c), (d), the boundary heat source is replaced by eighteen pairs of dipoles. (e), (f) show the temperature distribution along the line $y$=$0$, where the data assisted by the line boundary heat source (superimposed dipoles) are denoted as scatter points (solid line). The reference temperature distribution (black dashed line) is also plotted.}
	\label{fig6}
\end{figure}

The gradient ratio $\lambda$ measures how large a linear heat source needs to be introduced into a device under an external temperature field to offset external disturbances. The concentrating coefficient $\eta$ represents the ratio of the temperature gradient inside the core to the background temperature gradient. These two parameters provide a standard for local temperature control within the elliptical core-shell meta-device. Figs.~\ref{fig5}(c) and \ref{fig5}(d) display $\lambda$ and $\eta$ as functions of $\kappa_s/\kappa_c$ and $\kappa_b/\kappa_c$, keeping geometrical parameters constant. To achieve a thermal cloak as visualized in Fig.~\ref{fig5}(b), a thin insulating layer ($ \rho_i$) is incorporated between the core and the outer shell. This arrangement ensures the core isotherms remain undisturbed and the boundary heat source is devised to offset external temperature perturbations.

In simulations conducted using COMSOL Multiphysics, a confocal elliptical structure is defined with specific geometrical parameters: $c$=$2.165$~cm, $\rho_c$=$0.549$~cm, $\rho_i$=$0.689$~cm, $\rho_l$=$0.852$~cm, $\rho_s$=$1.06$~cm, and $\kappa_c$=$100$~W~m$^{-1}$~K$^{-1}$. The background's dimensions are set at $10\times10$~cm$^2$. Figs.~\ref{fig6}(a) and \ref{fig6}(b) illustrate the temperature profiles for thermal transparency and cloaking, respectively, with specific values of $\kappa_s/\kappa_c$=$0.2$, $\kappa_b/\kappa_c$=$0.5$, and $\nabla T_l$=$-712$~K~m$^{-1}$. In Figs.~\ref{fig6}(c) and \ref{fig6}(d), the line boundary heat sources are substituted with 18 pairs of dipoles, and temperature values are computed via $x \nabla T_l$. The temperature distributions along the central axis are also depicted in Figs.~\ref{fig6}(e) and \ref{fig6}(f). It is evident that temperatures generated by multi-dipole arrangements align well with those produced by the boundary heat source, underscoring the accuracy of this dipole superposition approximation.

\subsection{Spherical meta-devices resorting to diffusion-superposition dipoles}
The three-dimensional core-shell structure in a finite background can be described using spherical coordinates, and the heat conduction equation in this system can be written as:
\begin{equation}
\begin{aligned}
&\frac{1}{r^{2}} \frac{\partial}{\partial r}\left(r^{2} \frac{\partial T}{\partial r}\right)+\frac{1}{r^{2} \sin \theta} \frac{\partial}{\partial \theta}\left(\sin \theta \frac{\partial T}{\partial \theta}\right)\\
&+\frac{\partial^{2} T}{\partial \varphi^{2}} \frac{1}{r^{2} \sin ^{2} \theta}=0.    
\end{aligned}
\end{equation}

The temperature distribution in the core $T_{e c}$, shell $T_{e s}$, and background $T_{e b}$ can be written as:
\begin{equation}\label{Te3}
\left\{\begin{aligned}
T_{e,c}&=A_{c} r \cos \theta +T_0,\\
T_{e,s}&=A_{s} r \cos \theta+B_{s} r^{-2} \cos \theta+T_0, \\
T_{e,b}&=A_{b} r \cos \theta+B_{b} r^{-2} \cos \theta +T_0 .
\end{aligned}\right.
\end{equation}

In the presence of a thermal dipole in the shell of the core-shell structure, thermal fields emerge in the core and background regions. Similar to the two-dimensional case, the temperature distributions are approximated as: $T_d \approx M\cos \theta/(4\pi \kappa r^{2})$ for $r \gg L$, $T_{d} \approx A r \cos \theta+M\cos \theta /(4 \pi \kappa_{2} r^{2}) $ for $r \sim L$, and $T_d \approx 2 M r \cos \theta /(\pi \kappa L^{3})$ for $r \ll L$. 

Then, the total thermal fields in the core and background can be expressed as follows:
\begin{equation}\label{Tt3}
\left\{\begin{aligned}
T_{t,c}&=(\frac{2M}{\pi \kappa_c L^3}+A_c)r  \cos \theta+T_0,\\
T_{t,b}&=\nabla T_{u}r\cos \theta+(B_b+D_b)r^{-2}\cos \theta+T_0.
\end{aligned}\right.
\end{equation}
By choosing appropriate parameters, different functionalities of thermal metadevices can be realized. Similar to the two-dimensional case, dipole moment $M$ can be  adjusted to achieve thermal transparency in various background conductivity $\kappa_b$, where the key equation can be written as
\begin{equation}\label{trans3}
\left\{\begin{aligned}
B_b(\kappa_b)+D_b(\kappa_b,M)=0,\\
\frac{2M(\kappa_b)}{\pi \kappa_c L^3}+A_c(\kappa_b)=\eta \nabla T_u.
\end{aligned}\right.
\end{equation}

\begin{figure}
\centering
	\includegraphics[width=\linewidth]{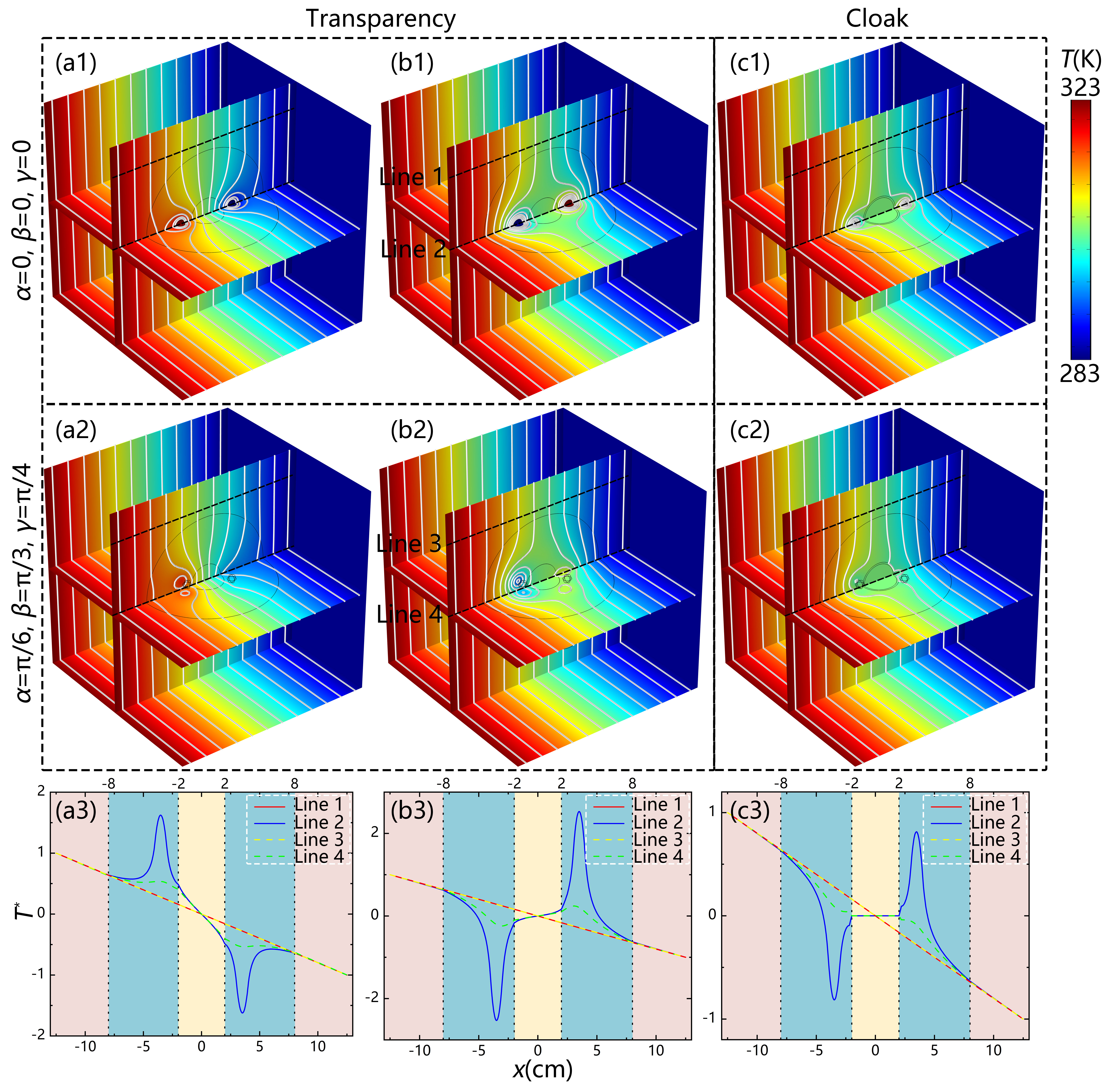}
    \caption{Temperature profiles of three-dimensional (a1), (b1) transparency, and (c1) cloak without rotation. (a2)-(c2) The temperature profiles of triple dipole-induced metadevices with rotating angles $\alpha$=$\pi/6$, $\beta$=$\pi/3$, and $\gamma$=$\pi/4$. (a3)-(c3) The central temperature distribution of (a1)-(c2).}
    \label{fig7}
\end{figure}
Solving Eq.~(\ref{trans3}) provides the specific values for $M$. Figs.~\ref{fig7}(a1)-\ref{fig7}(c1) display the temperature profiles of thermal transparency and cloak in the absence of rotation. Subsequent analyses focus on the core-shell structure subjected to an arbitrary external heat flux. This investigation involves rotating the finite background through a series of fixed-axis rotations ($x$-, $y$-, and $z$- axes). For simulation purposes, three pairs of dipoles were successively rotated by angle $\gamma$ around the $z$-axis, angle $\alpha$ around the $x$-axis, and angle $\beta$ around the $y$-axis. This sequence is tantamount to executing a reverse rotation on the background. The coordinates of the dipoles, both pre and post-rotation, adhere to the relation ${\bm r_i^{\prime}}$=$\bm Y(\beta) \bm X(\alpha) \bm Z(\gamma) {\bm r_i}$, where the index $i$ spans values from 1 to 3, denoting the three pairs of dipoles. The rotation matrices are given by,
\begin{equation}\label{rotation_matrix}
\begin{aligned}
 \bm X(\alpha)&=\left(\begin{array}{ccc}
  1 &0  &0 \\
  0 & \cos \alpha  &-\sin \alpha \\
  0 & \sin \alpha &\cos \alpha
\end{array}\right),\\
\bm Y(\beta)&=\left(\begin{array}{ccc}
  \cos \beta &0  &\sin \beta \\
  0 & 1  &0 \\
  -\sin \beta & 0 &\cos \beta
\end{array}\right),\\
\bm Z(\gamma)&=\left(\begin{array}{ccc}
  \cos \gamma &-\sin \gamma  &0 \\
  \sin \gamma & \cos \gamma  &0 \\
  0 & 0 &1
\end{array}\right)   
\end{aligned}
\end{equation}
Similar to the two-dimensional case, the component dipole moments are equal to the projections of the effective dipole moment on coordinate axes, namely, $M_1=M \bm r_1^{\prime} \cdot \bm e^{\bm x}/\left |  \bm r_1^{\prime} \right |$, $M_2=M \bm r_2^{\prime} \cdot \bm e^{\bm x}/\left |  \bm r_2^{\prime} \right | $, and $M_3=M \bm r_3^{\prime} \cdot \bm e^{\bm x}/\left |  \bm r_3^{\prime} \right |$. Figs.~\ref{fig7}(a2)-\ref{fig7}(c2) illustrate the temperature contours of the thermal transparency and cloak after rotating the angles $\alpha$=$\pi/6$, $\beta$=$\pi/4$, and $\gamma$=$\pi/3$. Moreover, the dimensionless temperature distributions along specific lines are presented in Figs.~\ref{fig7}(a3)-\ref{fig7}(c3). These results demonstrate that the metadevices perform well both before and after rotation, validating the scattering cancellation superimposed effect of the thermal dipole.

\subsection{Superimposed dipoles assisted thermotics extended to conduction-radiation field}
As temperature rises, thermal radiation becomes a factor that cannot be ignored. This necessitates modifications to the dipole-assisted thermotics. The Rosseland diffusion approximation~\cite{XuPRAP19a,XuPRAP20} is introduced to address the radiative model of participating media like aerogels. In a passive and steady heat transfer process, the total heat flux, denoted as \( \bm J_{\text{total}} \) and comprising the conductive flux \( \bm J_{\text{con}} \) and the radiative flux \( \bm J_{\text{rad}} \), is divergence-free:
\begin{equation}\label{rad}
\bm \nabla \cdot \left(\bm J_{\text{con}}+\bm  J_{\text{rad}}\right)=0
\end{equation}
The conductive flux is governed by the Fourier Law: $\bm J_{\text{con}}$=$-\kappa \bm\nabla T$ and the radiative flux is given by the Rosseland diffusion approximation: $\bm J_{\text{rad}}$=$-\gamma T^3 \bm\nabla T$. Here, $\gamma$=$16\tau^{-1}n^2\sigma/3$,  $\tau$ is the Rosseland mean extinction coefficient,
$n$ is the relative refractive index, and $\sigma$ is the Stefan-Boltzmann constant (5.67$\times$ 10$^{-8}$~W~m$^{-2}$~K$^{-4}$). Using these expressions, Eq.~(\ref{rad}) can be rewritten as:
\begin{equation}\label{rad2}
\bm \nabla \cdot \left[\left(\kappa+\gamma T^3\right)\bm \nabla T\right]=\nabla \cdot\left[\kappa \nabla\left(T+\frac{\gamma T^4}{4\kappa}\right)\right]=0
\end{equation}

\begin{figure}
\centering
	\includegraphics[width=\linewidth]{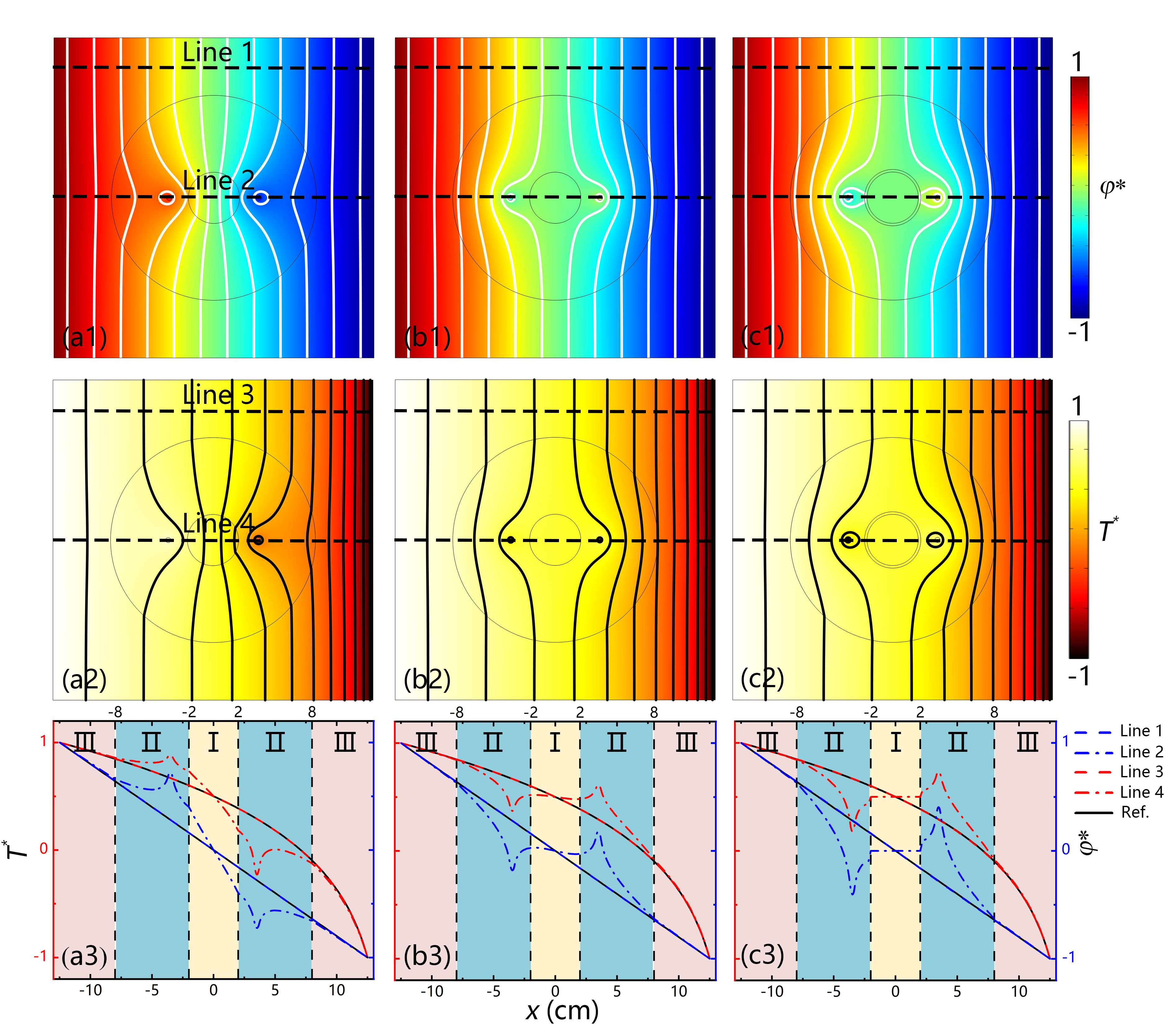}
	\caption{Simulative results of dipole-induced conduction-radiation metadevices. (a1)-(a3) Transparency with enhanced center. (b1)-(b3) Transparency with weakened center. (c1)-(c3) Cloak. The profiles of $\phi^*$ and $T^*$ are respectively presented in (a1)-(c1) and (a2)-(c2). Field distributions along lines 1-4 are also plotted.}
	\label{fig8}
\end{figure}

For the purpose of simplifying the conduction-radiation equation, the ratio \( \gamma/\kappa \) is considered constant across the entire domain. An auxiliary field \( \varphi = T^4 + \gamma T/(4\kappa) \) is introduced, enabling Eq.~(\ref{rad2}) to be restated as:
\begin{equation}
\nabla \cdot(\kappa \nabla \varphi)=0.
\end{equation}

Similar to the approach discussed above, the total distributions of \( \varphi \) in various regions are derived:
\begin{equation}\label{pt2}
\left\{\begin{aligned}
\varphi_{tc}&=(\frac{2M}{\pi \kappa_c L^2}+A_c)r \cos \theta+\varphi_0,\\
\varphi_{tb}&=A_{b} r\cos \theta+(B_b+D_b)r^{-1}\cos \theta+\varphi_0.
\end{aligned}\right.
\end{equation}
The undetermined coefficients can be calculated by matching conditions:

\begin{equation}
\left\{\begin{aligned}
A_{c}&=\frac{4 \nabla \varphi_{u} \kappa_{s} \kappa_{b} R_{s}^{2}}{\chi_1+\chi_2},\\
B_{b}&=\frac{-\nabla \varphi_{u}\left(\kappa_{c}-\kappa_{s}\right)\left(\kappa_{s}+\kappa_{b}\right) R_{c}^{2} R_{s}^{2}}{\chi_1+\chi_2}\\
&+\frac{\nabla \varphi_{u}\left(\kappa_{c}+\kappa_{s}\right)\left(\kappa_{s}-\kappa_{b}\right) R_{s}^{4}}{\chi_1+\chi_2}\\
D_{b}&=\frac{M}{\pi (\kappa_s+\kappa_b)},\\
A_b&=\nabla \varphi_{u},\\
\chi_1&=\left(\kappa_{c}-\kappa_{s}\right)\left(\kappa_{s}-\kappa_{b}\right) R_{c}^{2}\\
\chi_2&=\left(\kappa_{c}+\kappa_{s}\right)\left(\kappa_{s}+\kappa_{b}\right) R_{s}^{2}
\end{aligned}\right.
\end{equation}

The given equations provide a framework to achieve conduction-radiation transparency and cloak through the adjustment of \( \kappa \) and \( M \). The parameter $\gamma$ affects the radiative heat flux and the ratio $\gamma/\kappa$ is set as 3$\times$10$^{-8}$ in the whole system. To illustrate the concept of conduction-radiation transparency and cloak, finite-element simulations were executed within a square background possessing a side length \( L_0 = 25 \) cm and relative refractive index $n=1$. The left and right boundaries were maintained at \( T_L = 873 \) K and \( T_R = 273 \) K, respectively, while the upper and lower boundaries were deemed adiabatic. The gradient \( \nabla \varphi_{u} \) was determined using the formula \( [\varphi(T_H)-\varphi(T_L)]/L_0 \). Furthermore, dimensionless temperatures, \( T^* \) and \( \varphi^* \), were defined as \( (T-T_0)/(T_L-T_R) \) and \( (\varphi-\varphi_0)/(\varphi_L-\varphi_R) \), respectively, to enable a quantitative comparison of the field distributions across different thermal metadevices.

The simulation results are presented in Fig.~\ref{fig8}. Notably, the distributions of \( \varphi^* \) in Figs.~\ref{fig8}(a1)-\ref{fig8}(c1) appear analogous to those of \( T \) in the absence of thermal radiation. However, the temperature profiles of the conduction-radiation metadevice, depicted in Figs.~\ref{fig8}(a2)-\ref{fig8}(c2), manifest distinct characteristics. To gauge their efficacy quantitatively, the field distributions of \( T^* \) and \( \varphi^* \) along lines 1-4 are plotted in Figs.~\ref{fig8}(a3)-\ref{fig8}(c3). Upon comparing these profiles with the reference field distribution (black solid lines), it becomes evident that temperature profiles exterior to the shells remain unaltered, as though the core-shell structures were absent. Concurrently, the isotherms within the cores materialize as expected.

\section{Experimental results and discussion}\label{3}
\begin{figure*}
\centering
	\includegraphics[width=\linewidth]{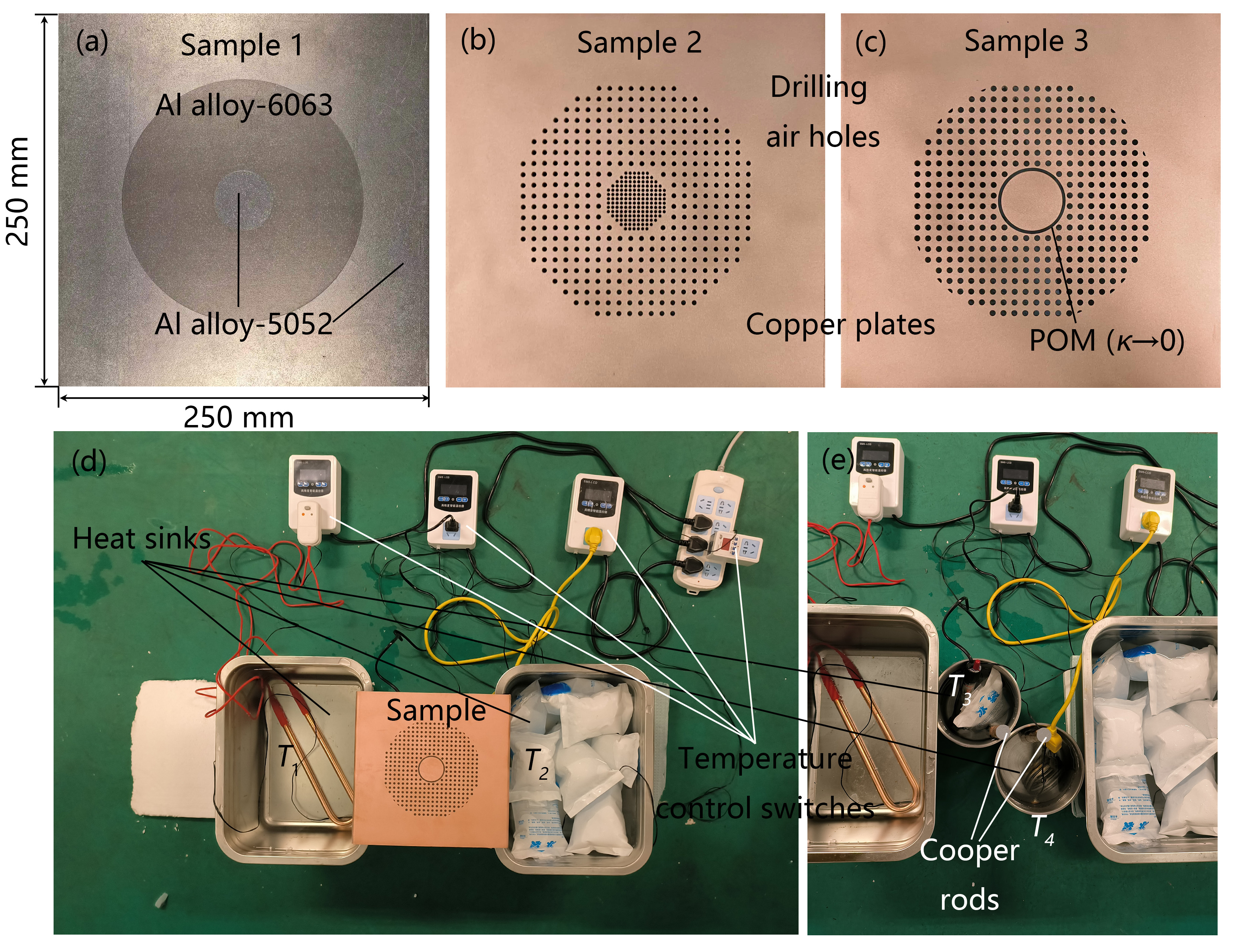}
	\caption{Fabricated samples and experimental setups. (a) Sample 1: enhanced-center transparency. (b) Sample 2: weakened-center transparency. (c) Samples 3: cloak. (d)(e) Photograph of the experimental setups.}
	\label{fig9}
\end{figure*}
\subsection{Experimental verification of diffusive superimposed dipoles}

\begin{table*}
\caption{The fabrication parameters of three samples. The effective thermal conductivities of different regions denote as $\kappa^{eff}$. Al: Aluminum; and POM: Polyformaldehyde.}\label{tab1}
\begin{ruledtabular}
\begin{tabular}{cccccc}
Region&Material&Drilling Radius (mm)&Drilling amount&$\kappa^{eff}$(W~m$^{-1}$~K$^{-1}$)\\ \hline
Core and background (1)& Al-5052&-&-&140\\
    Shell (1)&Al-6063&-&-&200\\
    Core (2)&Copper&0.9&156&200\\
	Shell (2)&Copper&1.5&360&304.1\\
    Background (2)&Copper&-&-&400\\
    Core and background (3)&Copper&-&-&400\\
	Shell(3)&Copper&2.6&212&250\\
    Insulated layer(3)&POM&-&-&0.06\\

\end{tabular}
\end{ruledtabular}
\end{table*}

\begin{figure*}
	\includegraphics[width=\linewidth]{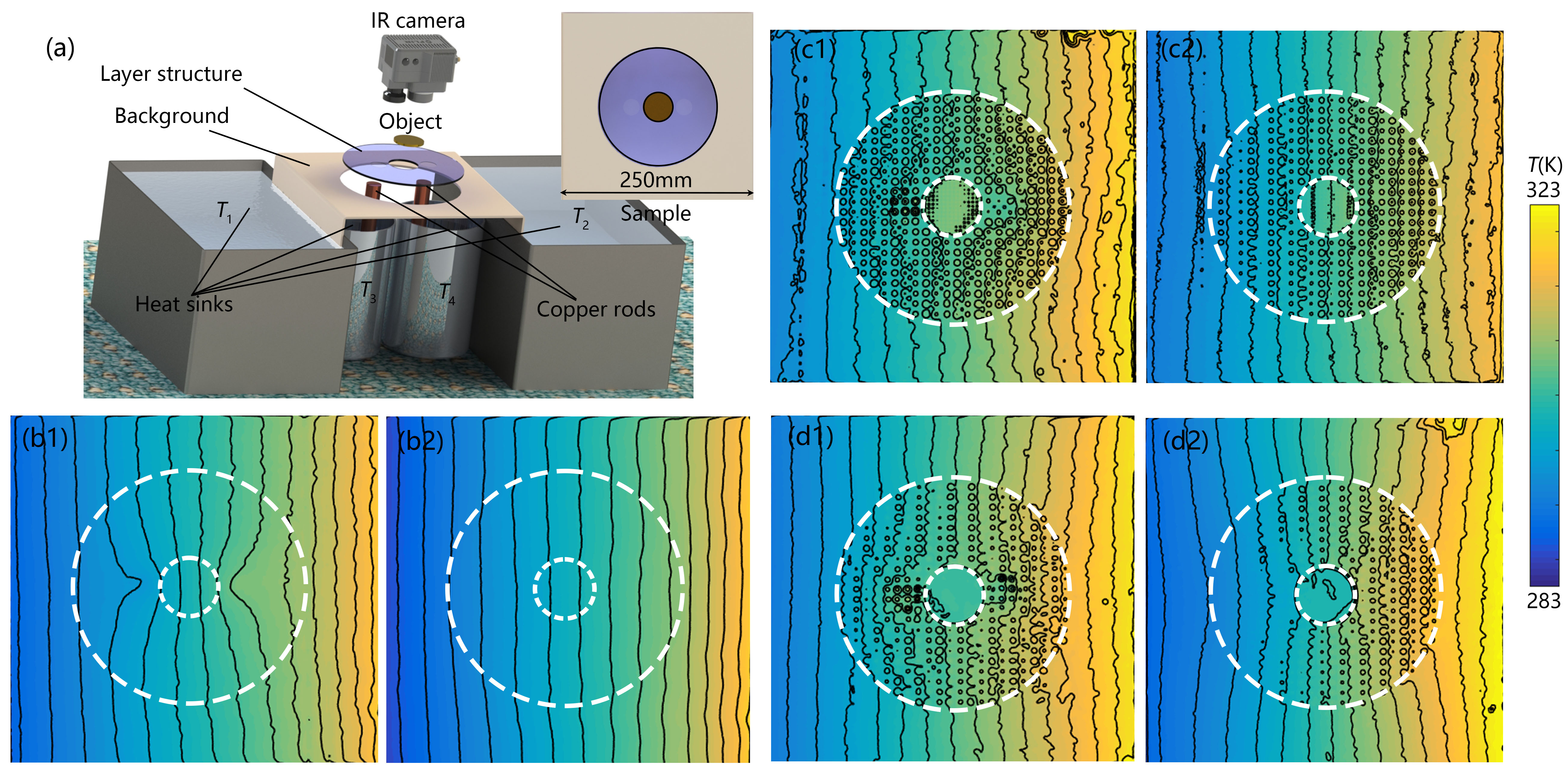}
	\caption{Experimental demonstrations of the proposed schemes and measured temperature fields. (a) Schematic of the experimental setup. Experimental measurements of sample 1 (b1) with a dipole and (b2) without a dipole, sample 2 (c1) with a dipole and (c2) without a dipole, and sample 3 (d1) with a dipole and (d2) without a dipole.}
	\label{fig10}
\end{figure*}

In order to further demonstrate the above derivations, three representative devices were manufactured based on the simulated structure: weakened-center transparency, enhanced-center transparency, and cloak, as shown in Figs.~\ref{fig9}(a)-\ref{fig9}(c). In all samples, the background size is \(25\times 25\)~cm\(^2\); the core and shell radii are \(R_c=2\)~cm and \(R_s=8\)~cm; and the dipole distance is \(L=7\)~cm. Two shallow slots with a radius of 1.1~cm were created in the bottom surface of each sample to ensure precise contact of the thermal dipole at the intended location. The upper (lower) surface was then covered with transparent plastic (insulating tape) to mitigate interference from infrared reflection (heat convection). Sample 1 was fabricated from aluminum alloy-6063 and aluminum alloy-5052 using high-pressure mechanical inlay technique. For samples 2 and 3, different region conductivities were achieved by drilling holes in a copper plate in line with the effective medium theory~\cite{YangAPL17a}. The specific parameters for each sample are detailed in Table~\ref{tab1}. Figs.~\ref{fig9}(d)-\ref{fig9}(e) display the experimental setups, where the external sources were simulated by submerging the lips into tanks filled with hot and cold water (\(T_1=323\)~K and \(T_2=283\)~K). Recognizing the challenge of embedding thermal dipoles into samples in actual experiments, the research team connected the bottom surface of the sample with two copper rods. These rods made contact with two additional heat baths (\(T_3\) and \(T_4\)), thereby emulating the temperature field of a dipole. The temperatures \(T_3\) and \(T_4\) were derived from finite element simulations and varied based on the radius of the heater (or cooler). Four heating rods (1~cm radius) equipped with temperature control switches were used to maintain a constant temperature in the heat baths. The lips of the sample were first submerged in the water tanks (\(T_1\) and \(T_2\)). Immediately thereafter, copper rods coated with thermally conductive silicone grease were attached to the sample's bottom [Fig.~\ref{fig10}(a)]. Approximately five minutes post-heating, a stable temperature distribution on the surface was observed and recorded using an infrared camera. This thermal imaging data underwent post-processing, which included extracting surface temperature values and drawing isotherms.

Figs.~\ref{fig10}(b1)-\ref{fig10}(d1) illustrate the temperature profiles and isotherms for enhanced-center transparency (\(\eta>1\)), weakened-center transparency (\(\eta<1\)), and cloak. The core isotherms are respectively concentrated, sparse, and absent altogether, with the background isotherms remaining linear. For reference, the temperature distribution with the thermal dipole deactivated is depicted in Figs.~\ref{fig10}(b2)-\ref{fig10}(d2). Without metadevices, the reference group's isotherms appear curved, showcasing the thermal dipole's pivotal role in heat distribution manipulation. The experimental results reveal that the metadevices' capability in temperature field regulation and in achieving the target thermal transparency and cloaking effects. Notably, the interface thermal resistance's (ITR) influence was an inherent aspect of the experiment. Nonetheless, this effect was counteracted by adjusting the water bath's temperature, tantamount to altering the dipole moment \(M\).

\subsection{The influence of contact thermal resistance}

\begin{figure}
\centering
	\includegraphics[width=\linewidth]{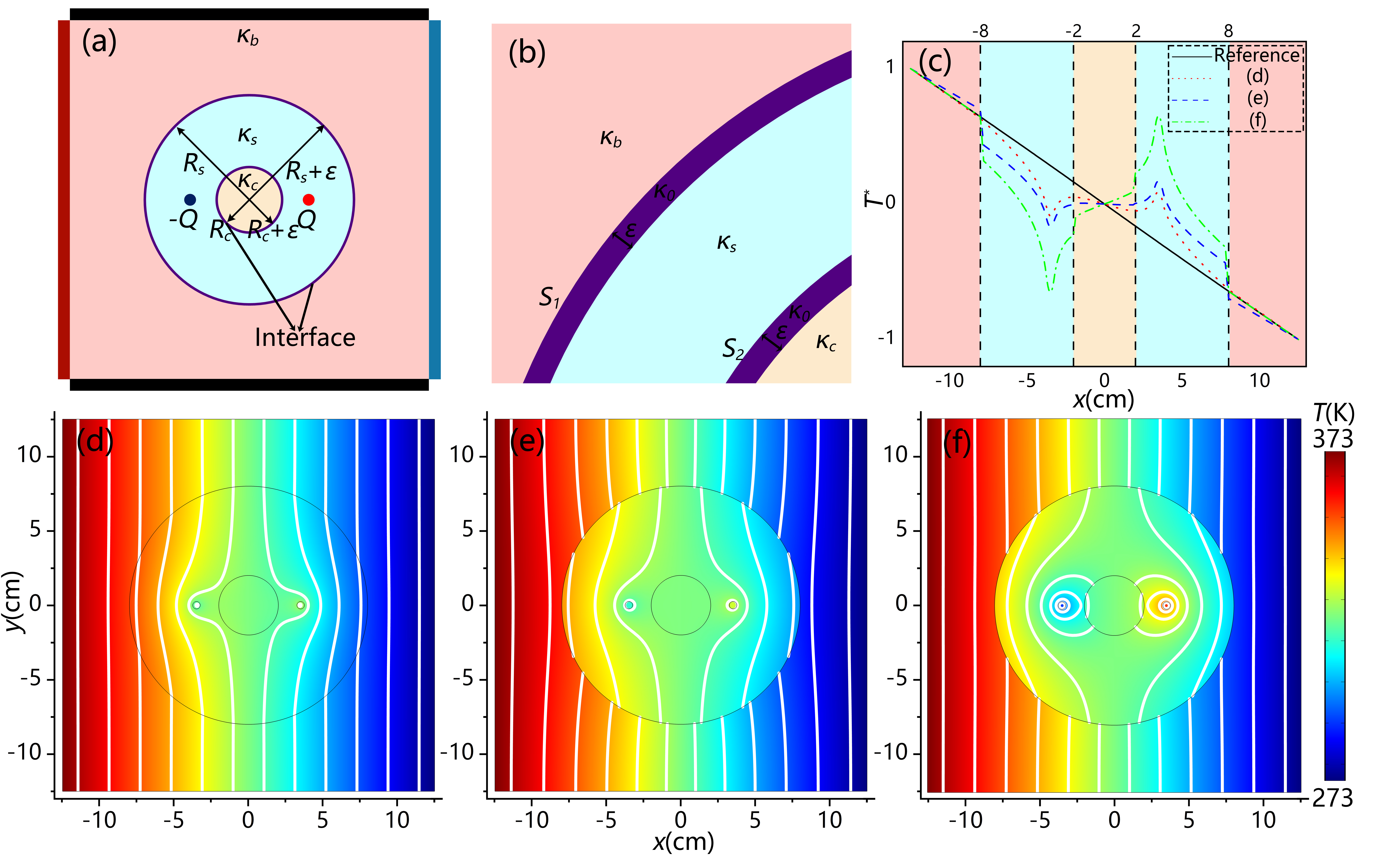}
	\caption{(a) Schematic diagram of dipole-assisted metadevice with ITR. (b) Schematic of ITR. (c) Dimensionless temperature distribution $T^*$ along central axes of panels (d), (e), and (f). The temperature profiles of (d) without ITR, (e) with ITR, and (f) with ITR and enhanced power dipole.}
	\label{fig11}
\end{figure}

In previous analyses, the effect of ITR at the interface was set aside to elucidate underlying physics. Nonetheless, the presence of ITR at the junction of two materials makes it a crucial aspect worthy of detailed exploration. Drawing an analogy to the concept of electrical resistance, the thermal resistance can be expressed as~\cite{ChenRMP22}:
\begin{equation}\label{ITR}
R_t=\frac{\Delta   T}{J},
\end{equation}
where \( \Delta T \) denotes the temperature discontinuity at the interface, and \( J \) represents the heat flux density. Researchers can employ this formula to experimentally measure ITR values when disparate materials come into contact. Prior studies have indicated a positive correlation of ITR with surface roughness and interface thickness, while noting an inverse relation with thermal conductivity of interphase. As a result, methods such as polishing, the application of high-pressure mechanical inlay technique, and the usage of thermal grease have been identified to significantly mitigate ITR. At the microscopic level, when entities such as two carbon nanotubes come into contact, the ITR has been observed to rely heavily on the contact area. In contrast, for macroscopic interfaces, ITR remains unaffected by changes in the interface area, indicating its geometric independence.

The ITR for two dry-contact copper pieces is approximately 3$\times$10$^{-4}$~K~m$^2$~W$^{-1}$. When utilizing high-pressure mechanical inlay technique, as demonstrated in experimental sample 1, there is a notable reduction in ITR. Conservatively, based on the structure shown in Fig.~\ref{fig2}(e), we consider the ITR of both inner and outer boundaries of the shell to be 1$\times$10$^{-4}$~K~m$^2$~W$^{-1}$. Figs.~\ref{fig11}(a) and \ref{fig11}(b) show the overarching structural framework and the schematic of the interface, respectively, with an exaggerated thickness of the interface. Fig.~\ref{fig11}(d) displays the temperature distribution without considering contact resistance, and it is identical to the temperature distribution in Fig.~\ref{fig2}(e), the ITR for both the internal and external boundaries is set as 1$\times$10$^{-4}$~K~m$^2$~W$^{-1}$. The ensuing temperature profile, depicted in Fig.~\ref{fig11}(e), clearly evidences perturbations in the background. The isotherms at the boundaries of the shell display noticeable discontinuities. This can be attributed to the shell's contact with minuscule air layers at its boundaries, resulting in diminished effective thermal conductivity. Drawing from observations in Fig.~\ref{fig2}(b), a shell with reduced thermal conductivity demands a larger thermal dipole moment to nullify the background thermal perturbations. As shown in Fig.~\ref{fig11}(f), the dipole moment is amplified to 2.1 times that in Fig.~\ref{fig11}(e), and the background disturbance was largely counteracted. Fig.~\ref{fig11}(c) offers a visual representation of the temperature distribution along the axis in panels (d), (e), and (f). A careful comparison reveals that the background disturbance resulting from introducing ITR is offset by the heat source with an increased power dipole. Such observations underscore the method's inherent robustness and adaptability in practical applications.

\section{Conclusion}\label{4}

In summary, a theory is proposed for active scattering cancellation in the Laplace field using a dipole heat source. Leveraging the combined influence of the far-field and near-field from superimposed thermal dipoles, two innovative thermal meta-devices have been conceptualized and designed, including transparency and cloak. These meta-devices can regulate the temperature gradient in localized regions by adjusting the magnitude of the dipole moment, without disturbing the ambient temperature field. The demonstrations reveal that the proposed scheme significantly exhibits local temperature manipulation and adapt to different environments without the need to alter inherent parameters (positions of superimposed dipoles and configuration of thermal conductivity), and they are still robust even after considering the interface thermal resistance. Our schemes can be implemented across 2D and 3D cases, as well as in geometrically isotropic and anisotropic scenarios using isotropic and homogeneous materials. Compared to previous active thermal control methods, the superimposed dipole control strategy offers a novel approach for the simple and efficient management of far and near thermal fields. Due to the similarity between the governing equations of the direct current electrical model~\cite{YangAM15,HuangPRE03} or the simplified hydrodynamics model~\cite{DaiPRE23,GaoJPCC07,QiuJPCB15} and the heat conduction equation, the principle of superimposed dipole-assisted regulation can be widely applied to these fields. Moreover, these designs are anticipated to be applied in the thermal management of precision structural components and offer a novel approach for the automated manipulation of heat transfer using machine learning. 

\section*{AUTHOR DECLARATIONS}
\subsection*{Conflict of Interest}
The authors have no conflicts to disclose.

\section*{Author Contributions}
{\bf Pengfei Zhuang}: Conceptualization, Investigation, Methodology, Validation, Software, Visualization, Writing-original draft. {\bf Xinchen Zhou}: Conceptualization, Investigation, Writing-original draft. {\bf Liujun Xu}: Conceptualization, Investigation, Supervision, Writing-original draft. {\bf Jiping Huang}: Conceptualization, Project administration, Supervision, Writing-review \& editing.

\section*{DATA AVAILABILITY}
The data that support the findings of this study are available from the corresponding author upon reasonable request.

\section*{ACKNOWLEDGMENTS}
J. H. acknowledges the financial support provided by the National Natural Science Foundation of China under Grants No. 12035004 and No. 12320101004, the Science and Technology Commission of Shanghai Municipality under Grant No. 20JC1414700, and the Innovation Program of Shanghai Municipal Education Commission under Grant No. 2023ZKZD06. L. X. is supported by the National Natural Science Foundation of China (12375040, 12088101) and NSAF (U2330401).

\bibliography{cas-refs}

\end{document}